\documentclass[preprint,12pt]{elsarticle}

\usepackage{amssymb}
\usepackage{graphicx} 
\usepackage{xcolor}
\usepackage{float}
\usepackage[version=4]{mhchem}
\usepackage{svg}
\usepackage{ulem}

\newif\ifsubmit
\submittrue

\ifsubmit
    \newcommand{\can}[1]{}
    \newcommand{\todo}[1]{}
    \newcommand{\tocite}[1]{}
\else
    \definecolor{comments}{rgb}{0.1, 0.66, 0.1}
    \newcommand{\can}[1]{[{\color{comments}CL: #1}]}
    \newcommand{\todo}[1]{[{\color{red}TODO: #1}]}
    \newcommand{\tocite}[1]{[{\color{red}citation-#1}]}
\fi

\journal{COSSMS}

\begin{document}

\begin{frontmatter}



\title{Current Opinions on Memristor-Accelerated Machine Learning Hardware}


\author[1]{Mingrui Jiang}
\author[2]{Yichun Xu}
\author[1]{Zefan Li}
\author[1,3,4]{Can Li}

\affiliation[1]{organization={Department of Electrical and Electronic Engineering, The University of Hong Kong},
            city={Pokfulam},
            state={Hong Kong SAR},
            country={China}}

\affiliation[2]{organization={School of Integrated Circuits, Huazhong University of Science and Technology},
            city={Wuhan},
            state={Hubei},
            country={China}}

\affiliation[3]{organization={Institute of Mind, The University of Hong Kong},
            city={Pokfulam},
            state={Hong Kong SAR},
            country={China}}

\affiliation[4]{organization={State Key Laboratory of Brain and Cognitive Sciences, The University of Hong Kong},
            city={Pokfulam},
            state={Hong Kong SAR},
            country={China}}
    
\begin{abstract}
The unprecedented advancement of artificial intelligence has placed immense demands on computing hardware, but traditional silicon-based semiconductor technologies are approaching their physical and economic limit, prompting the exploration of novel computing paradigms. 
Memristor offers a promising solution, enabling in-memory analog computation and massive parallelism, which leads to low latency and power consumption. 
This manuscript reviews the current status of memristor-based machine learning accelerators, highlighting the milestones achieved in developing prototype chips, that not only accelerate neural networks inference but also tackle other machine learning tasks. 
More importantly, it discusses our opinion on current key challenges that remain in this field, such as device variation, the need for efficient peripheral circuitry, and systematic co-design and optimization. 
We also share our perspective on potential future directions, some of which address existing challenges while others explore untouched territories. 
By addressing these challenges through interdisciplinary efforts spanning device engineering, circuit design, and systems architecture, memristor-based accelerators could significantly advance the capabilities of AI hardware, particularly for edge applications where power efficiency is paramount.


\end{abstract}

\begin{graphicalabstract}
\includegraphics[width=1\textwidth]{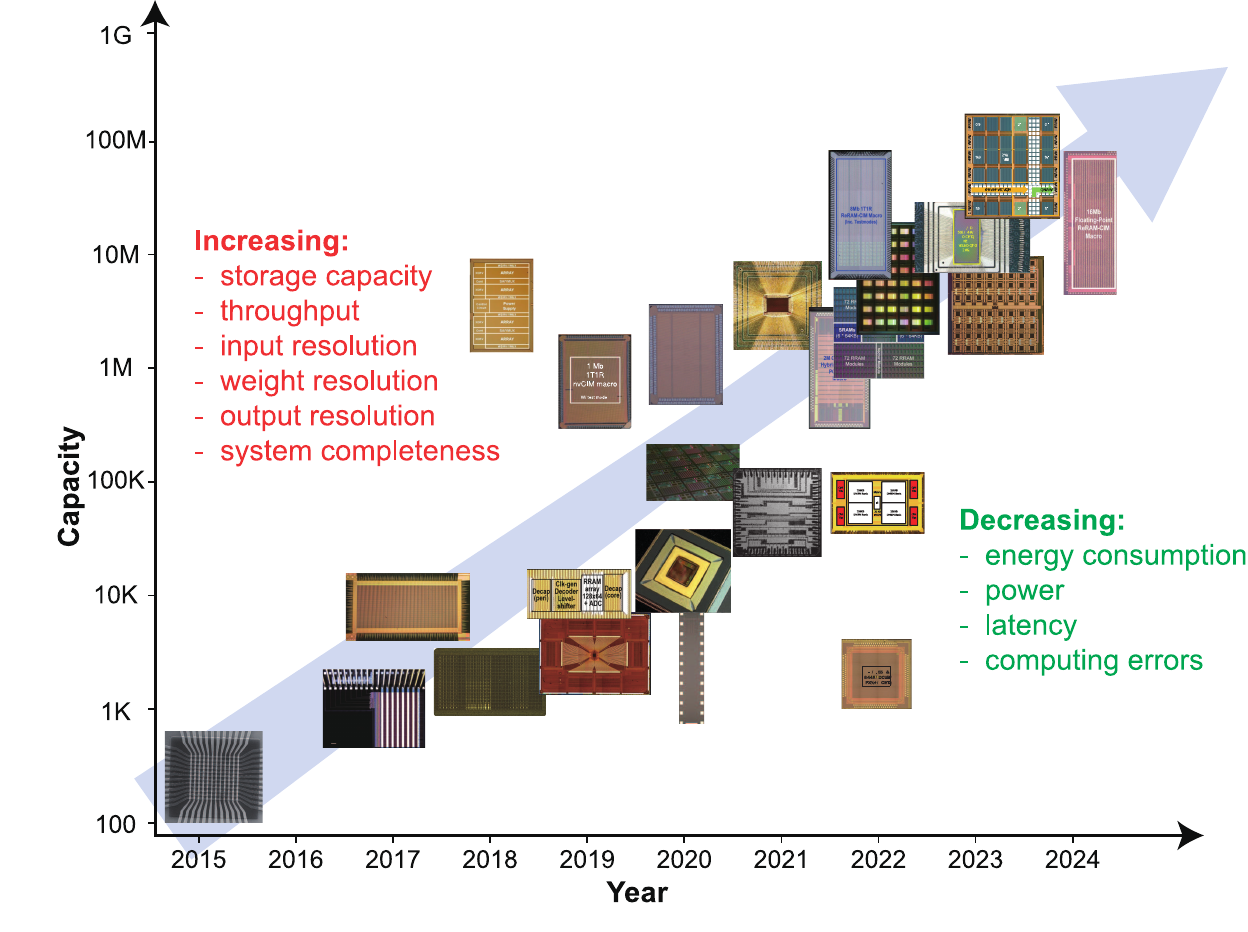}
\end{graphicalabstract}



\end{frontmatter}


\section{Introduction}

With the emergence of generative AI tools like chatGPT, artificial intelligence (AI) has revolutionized our everyday lives, outperforming humans in numerous fields. Yet, as these potent AI models grow more intricate, it's widely agreed that hardware performance is the bottleneck in their continued evolution. AI accelerator chips, such as those based on the general-purpose graphics processing unit (GPGPU), the field-programmable gate array (FPGA), and the application-specific integrated circuit (ASIC), are subjects of intensive research and development. The AI chip market value is estimated at US\$51.9 billion in 2023, while it is projected to surpass US\$1114 billion by 2032\cite{DHR2024aichip}. 

Over the past few decades, the progression of computing hardware has been driven by the scaling of silicon transistors, the fundamental building block of current integrated circuits. However, this trend has been slowing down recently as transistors near their physical limits. The rapidly increased demand for computational power, fueled by AI and Big Data analytics advances, is creating a significant gap between supply and demand. Consequently, it has become crucial to investigate new computing paradigms rooted in unconventional devices.

One example is to use resistive networks to perform matrix multiplication operations, which are the most time-consuming part of many AI algorithms.
Resistive networks for matrix multiplication operations and neural networks are not new. As early as 1960, researchers built a single neural network ``ADALINE''\cite{widrow1960adaptive} with physical variable resistors. But such a system with discrete components cannot scale. The introduction of the very large-scale integration (VLSI) later made it possible to fabricate smaller and thus more devices on a single chip to allow larger resistive networks. In 1989, a group of researchers from Bell Labs fabricated a 22$\times$22 resistive crossbar array for hardware neural networks (Hopfield)\cite{verleysen1989analog} . Still, the device representing synaptic weights was not re-programmable owing to the lack of available devices that could exhibit tunable resistance. The resistive switching memory devices, on the other hand, can provide such characteristics, and therefore have recently received much renewed interest. This trend was accelerated after Hewlett Packard Labs linked the device behavior to a postulated `memristor' concept in 2008\cite{chua1971memristor} , and this device demonstrated well tunable analog resistance behavior\cite{strukov2008missing} .

Memristors, or resistance switches, are emerging as analogue resistive memory. They have demonstrated great potential in accelerating existing workloads and developing new computing paradigms. The device is a two-terminal passive device with a resistive switching layer sandwiched between two electrodes. The current-voltage relationship of the device exhibits rich nonlinear behavior, enabling various processing capabilities based on physics. The switching mechanism of memristors is usually different from that of conventional silicon transistors, resulting in promising figures of merit such as scalability, switching speed, data retention, three-dimensional stackability, and more. Consequently, they are used as next-generation memory systems (e.g., ReRAM) that support large data sizes stored in on-chip memory and provide wide accessing bandwidth. Importantly, they can process information directly within the memory itself in a massively parallel manner, eliminating expensive data movement. 
As a result, computing hardware based on non-volatile resistive switches is projected to achieve at least a $100\times$ gain over CPU/GPU and at least a $10\times$ gain over SRAM-based ASIC accelerators in terms of performance density (in GOPS mm$^{-2}$) and power efficiency (in GOPS W$^{-1}$)\cite{wang2020resistive}.

This work aims to share our views on developing memristor-based machine learning accelerators, covering the current status, remaining challenges, and possible future directions. Some future directions involve addressing the remaining challenges, so the sections are not strictly separate. The challenges section also discusses efforts to tackle these issues, which could be seen as future directions.

\section{Current Status}

\subsection{Prototype chips for neural network inference}

In recent years, several research groups have independently developed prototypes and macro chips using silicon transistor-based peripheral circuits and monolithically integrated memristors for accelerating mostly the vector-matrix multiplication (VMM) operation for artificial neural network (ANN) inference.
This is because VMM is the most time- and energy-consuming step in ANNs with conventional von Neumann systemd digital hardware, because of their series processing and separated memory and compute units.
Memristive crossbars represent have emerged as a promising solution, demonstrating orders of magnitude advantages in both latency and energy efficiency. 
By arranging memristive devices to a crossbar structure, the matrix elements can be mapped to tunable cell conductance.
The input vectors can be encoded in different forms, including voltage amplitude, pulse widths, pulse numbers, or pulse intervals.
Using Ohm's law for multiplication and Kirchhoff's current law for accumulation, the dense crossbar can perform multiplication-accumulation (MAC) operations fully in parallel, with the computation occurring where the data was stored. 

\begin{figure}[h]
    \centering
    \includegraphics[width=1\textwidth]{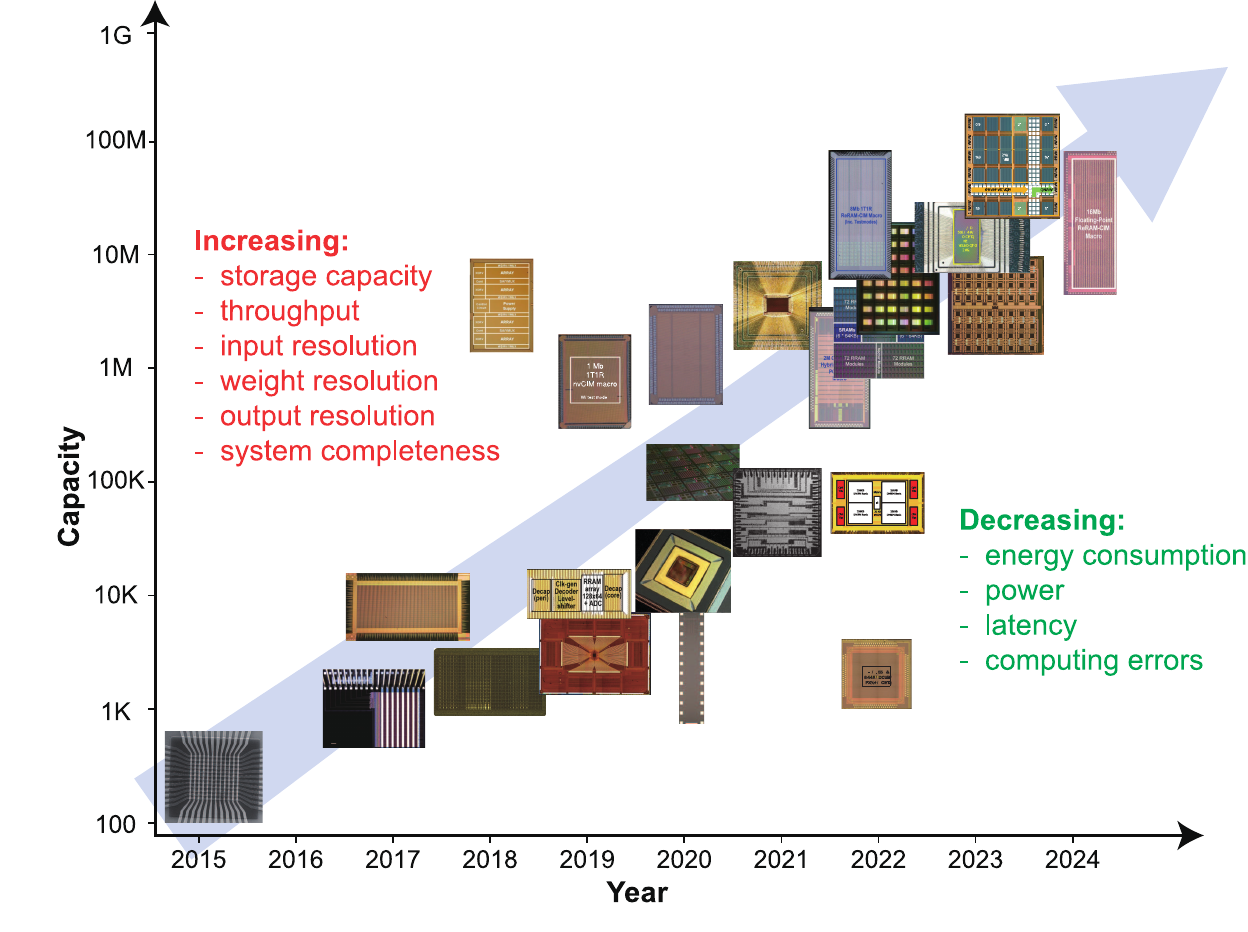}
    \caption{Exponential scaling trend for memristor-based VMM accelerator.}
    \label{fig:scaling}
\end{figure}

Fig. \ref{fig:scaling} summarizes recent reports that show the capacity of those prototype chips is increasing by about one order of magnitude each year, far faster than Moore's Law. Meanwhile, the computing resolution and system sophistication are continuously improving. More and more circuit components and functions needed in the ML tasks are integrated on chip demonstrating the end-to-end processing capability. The following paragraphs summarize key achievements in chronological order.
\begin{itemize}
\item
In 2015, researchers from UCSB pioneered a demonstration of a 12$\times$12 passive crossbar array with the ability of doing a simple pattern classification task with in-situ training.
\item
In 2017, researchers from Tsinghua experimentally demonstrated face classification task on a 128$\times$8 memristor crossbar array\cite{yao2017face}. 
In this demonstration, pulse number represented the input, and the current accumulation was assisted by software. 
In the same year, another group in Tsinghua collaborating with NTHU reported a process-in-memory chip with four 32$\times$32 1T1R RRAM arrays implementing binary input, binary weight, and binary output MVM\cite{su2017462gops}. 
In the following years, researchers from NTHU consistently worked to increase the array size, input precision, weight bit precision, output precision, and overall speed and energy efficiency. 
Their latest work in 2024 showed a 16Mb RRAM macro enabling floating-point computing with 31.2TFLOPS/W energy-efficiency\cite{wen202422nm}.
Meanwhile, in 2017, researchers from UMass Amherst and Hewlett Packard Labs, for the first time, demonstrated analog input, analog weight, and analog output on an integrated 128$\times$64 1T1M array with discrete off-chip peripheral circuits for analog signal processing and image compressing tasks\cite{li2017analogue}. 
Soon after, various ML algorithms have been experimentally implemented on the same platform, including in-situ training of multilayer perceptron (MLP)\cite{li2018efficient}, convolutional neural network (CNN)\cite{wang2019insitu}, long short-term memory (LSTM)\cite{li2019long} and reinforcement learning(RL)\cite{wang2019reinforcement}, etc.
\item
In 2018, Panasonic showcased a 4M storage-capacity memristor chip integrated with necessary peripherals implementing binary input, analog weight, and binary output MVM\cite{mochida20184m}. 
Similarly, researchers from Aix Marseille University demonstrated a 2k memristor chip with 2T2R differential cell structure and decoders and sense amplifiers integrated aiming for binary neural network\cite{bocquet2018memory}. 
\item
In 2019, UMich presented a fully integrated memristor chip consisting of on-chip DACs, ADCs, a controlling processor, and a 54$\times$128 passive array without transistor selector\cite{cai2019fully}.
In this implementation, the pulse width is used for representing the input associated with a current integrating ADC to read out the result.
In the same year, ASU/GaTech presented a binary input (1 or -1), binary weight (1 or -1), and 3b output VMM chip with a 128$\times$64 array\cite{yin2019monolithically}\cite{yin2020high}.
\item
In 2020, Tsinghua implemented a full hardware CNN with eight 128$\times$16 arrays and off-chip peripherals\cite{yao2020fully}. 
In the same year, they presented another chip integrated with 158.8kb memristors configuring as differential 2T2R arrays representing signed weight and resolution adjustable LPAR-ADCs\cite{liu2020fully}. 
Later, various edge learning tasks have been demonstrated on this chip including motion control, image classification, and speech recognition\cite{zhang2023edge}. 
In the meanwhile, researchers from Hewlett Packard labs presented an integrated memristor chip consisting of three 64$\times$64 arrays and on-chip drivers, TIAs, Multiplexers, and ADCs\cite{li2020cmos}. 
On this hardware platform, researchers have demonstrated the Hopfield neural network for NP-hard problem solving\cite{cai2020power}\cite{jiang2022efficient}\cite{jiang2023efficient}, bio-plausible training algorithms\cite{yi2022activity}, graph neural networks\cite{mao2023reram} and etc. 
\item
In 2021, IBM showed their HERMES core based on phase change memristors with a 14nm technology node. This core contains 256$\times$256 8T4R arrays, PWM modulators, and CCO-based ADCs enabling fully parallel 8-bit digital inputs and outputs \cite{khaddam2021hermes}\cite{khaddam2022hermes}. 
Moreover, in the core, a local digital processing unit was implemented for calibrations and ReLU activation.
\item
In 2022, the Chinese Academy of Science reported a VMM acceleration marco with 2kb 3D stacked memristors showing enhanced memory density\cite{huo2022computing}.
Later, Gatech reported a PWM input and ADC-free macro using PWM signal communication between the sub-arrays\cite{jiang202240nm}.
In the same year, researchers from Stanford presented a fully integrated 48-core chip aiming for versatile AI tasks, with each core containing a 256$\times$256 memristor array, voltage mode sensing, and transposable computing configurations\cite{wan2022compute}.
\item
In 2023, IBM improved their HERMES project chip to have 64 previously reported cores\cite{le202364core}.
In this chip, except from analog VMM, additional digital data processing needed for CNN and LSTM and inter-tile communications were integrated and demonstrated.
At the same time, IBM reported another chip that had 34 tiles, with each tile being capable of storing 512$\times$512 weights (4 PCMs for each weight)\cite{ambrogio2023analog}.
Moreover, a PWM signal was utilized for efficient communication between each tile which reduces the cost needed for data conversion.
\end{itemize}
In summary, the development of memristor-based VMM accelerator macros has been very rapid. 
Overall, it is progressing towards higher input, weight, and output resolution, larger arrays and higher memory density, more advanced technology nodes, more computing cores, and a more sophisticated system integrating all necessary processing blocks needed for complete AI inference/training tasks. 

\begin{figure}[H]
    \centering
    \includegraphics[width=1\textwidth]{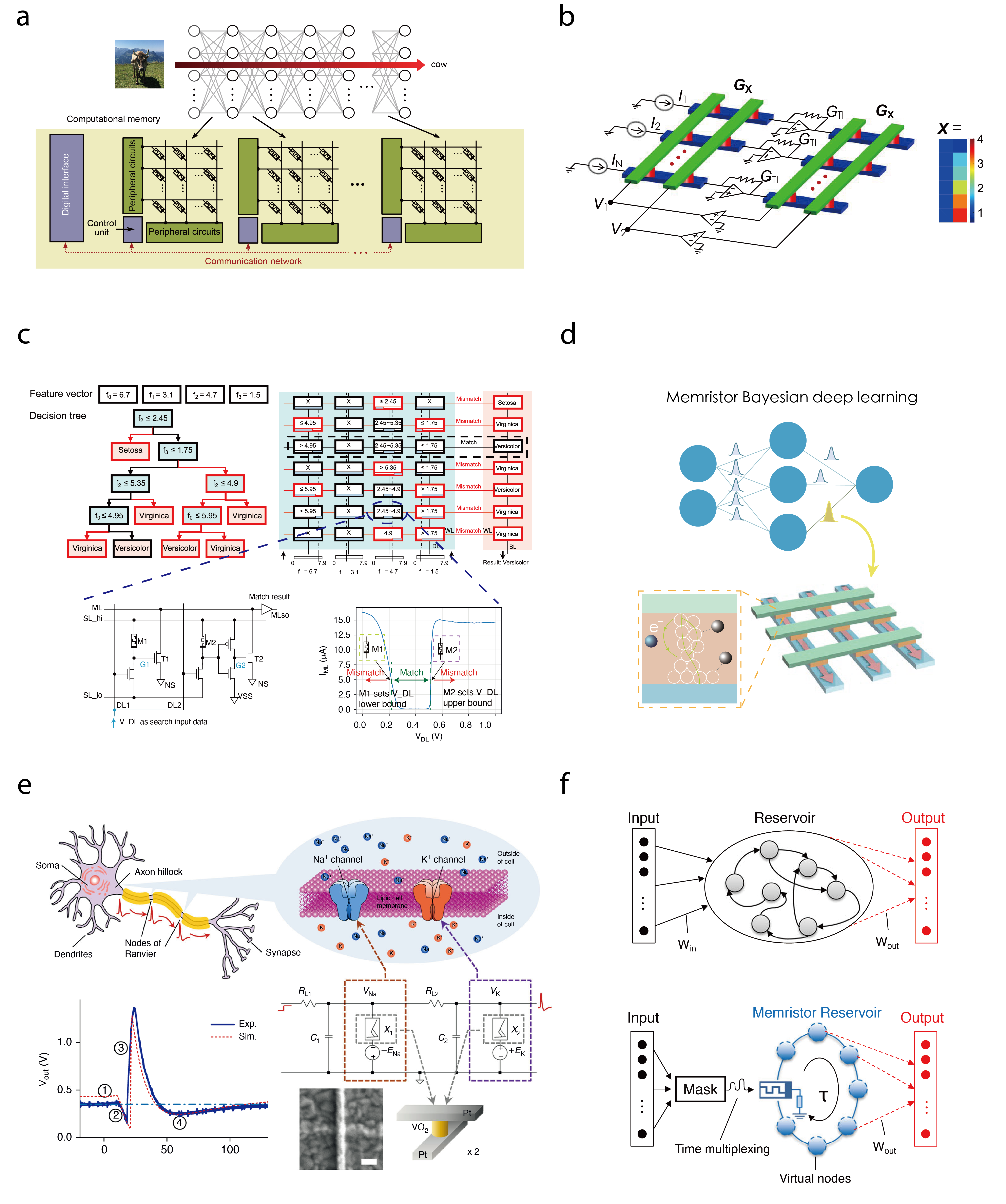}
    \caption{Various machine learning tasks can be accelerated by memristor. (a)Deep neural network, adapted from \cite{sebastian2020memory} (b)Linear regression, adapted from \cite{sun2020onestep} (c)Decision tree, adapted from \cite{pedretti2021treebased} (d)Bayesian neural network, adapted from \cite{lin2023uncertainty} (e)spiking neural network adapted from \cite{yi2018biological} (f)reservoir computing, adapted from \cite{zhong2021dynamic}}
    \label{fig:MLtasks}
\end{figure}
\subsection{Novel paradigms for various machine learning models}

Except for showing great potential as an VMM computing core for neural network acceleration, researchers were also actively exploring accelerating other machine learning models with varying in-memory analog operations using memristors. 

One example is utilizing the dynamic evolution process in recurrent-connected memristor crossbar structure to compute in the deep analog domain. 
With this structure, researchers have experimentally demonstrated one-step linear equation solving, matrix inversion, eigenvectors computing, and linear regression and classification\cite{sun2019solving}\cite{sun2020onestep}.  This shows great efficiency compared to conventional clocked iterative methods.

In addition to the crossbar structure which performs in-memory multiplication, researchers also explored memristor-based content addressable memory (CAM), which performs comparison operations for massively parallel in-memory look-up or key-value pairs searching. 
Various applications have been reported, including finite state machines, tree-based machine learning models, data mining, and reconfigurable computing\cite{karam2015emerging}. Recently, memristor-based ternary CAM (TCAM) has been experimentally implemented and applied for network security and genomics sequencing, showing great improvements in both area and energy consumption compared to conventional SRAM-based CAM\cite{graves2019memristor}\cite{graves2020inmemory}. 
An analog CAM that can fully utilize the analog conductance states of memristors has also been designed and its analog-searching functionality proved experimentally\cite{li2020analogcam}. 
This is further applied to random forests for accelerating interpretable AI models\cite{pedretti2021treebased}. 
Moreover, such CAM structures can be embedded into DNN models, enabling life-long on-device learning\cite{mao2022experimentally}.

Another intensively researched direction is utilizing the unique characteristics of different memristor devices, particularly their intrinsic stochasticity with rich non-linearity. 
Intrinsic stochasticity has been explored for implementing Bayesian probabilistic computing\cite{dalgaty2021insitu}\cite{dutta2022neural}\cite{li2022enabling}\cite{woo2022probabilistic}\cite{zheng2022hardware}\cite{lin2023uncertainty}, solving combinatorial optimization problems\cite{mahmoodi2019versatile}\cite{cai2020power}, conducting locality-sensitive hashing\cite{mao2022experimentally}\cite{yang2023self} and working as echo state machine\cite{wang2023echo}. 
The rich non-linearity including the dynamics of different types of memristors (drift and diffusive), has been utilized for reservoir computing\cite{du2017reservoir}\cite{moon2019temporal}\cite{zhu2020memristor}\cite{jang2021time}\cite{zhong2021dynamic}\cite{milano2022inmateria}\cite{zhong2022memristor} and bio-mimicking neuromorphic computing\cite{pickett2013scalable}\cite{wang2017memristors}\cite{wang2018fully}\cite{yi2018biological}\cite{yoon2018artificial}\cite{zhang2020artificial}\cite{zhang2021neuromorphic}. 
Recently, the unique I-V characteristics of molecular memristors have also been utilized for efficiently implementing decision-tree\cite{goswami2021decision}. 
We believe more features of memristors will be explored and utilized for developing novel computing paradigms and systems in the future, further expanding the application scenarios of memristor-based machine learning systems.

\section{Remaining Challenges}

\begin{figure}[h]
    \centering
    \includegraphics[width=1\textwidth]{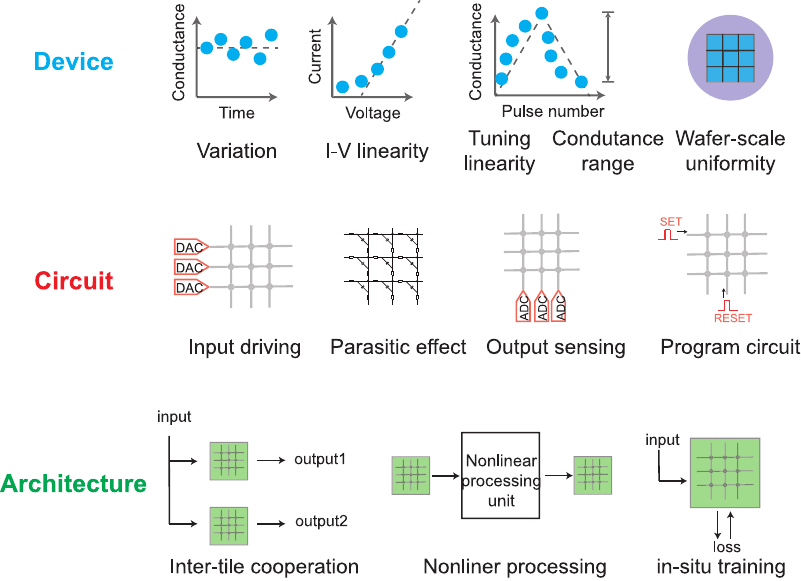}
    \caption{Challenges remaining for memristor-based ML accelerators.}
    \label{fig:challenge}
\end{figure}

Despite the exponential growth of the memristor-based machine learning accelerator prototypes at a faster face than the Moore's Law, they have not yet achieved commercial success. 
GPU will still be the dominate hardware for general-purposed machine learning hardware in the foreseeable future, while the memristor-based accelerator stand a better chance in edge device for a specialized application where the access to energy is highly limited. 
There are still challenges to address for commercial adoption, while addressing challenges might also provide opportunities for novel computing paradigms, which requires efforts across disciplines, from the device, circuit, system, to algorithms. 
At the device level, efforts focus on improving the device electrical performance, circuit level focus on the peripheral circuits of a single computing core, and system level focuses on inter-tile cooperation and communication and software hardware co-design.
\ref{fig:challenge} illustrates some of the challenges at different levels. 
It is noteworthy that addressing challenges at one level often requires knowledge at other levels, so cross-layer expertise is highly preferred in designing the memristor machine learning accelerators. 
Different machine learning models and application may impose highly diverse requirements on device, circuit and system designs, and thus poses very different challenges. Challenges in one scenario can even mean opportunities in another. 
For example, the large noise of memristor device is a troublesome thing for precise computing but can be a perfect randomness source for probabilistic computing. 
Therefore, it is essential to treat the challenges of current memristor-based ML accelerators dialectically.

\subsection{Device}

\subsubsection*{\textbullet{Device variation}}

One of the most prominent challenges for memristor-based accelerators is inaccurate computing due to device variation. In-memory computing accelerators store values in memory cells, but if the conductance values deviate from their targets, the results become inaccurate, leading to incorrect machine learning outcomes in the hardware. Conductance variation includes inaccurate programming, short-term conductance fluctuation, and long-term conductance drift.


Memristors' conductance states expectation can be predicted and simulated based on physical models \cite{yeAEM2022}.
However, conductance variations arise from stochastic physical processes. 
The conduction filament in memristor devices is typically just a few nanometers in size, so even a single atom fluctuation can significantly impact device conductance. This characteristic offers excellent scalability down to the nanometer scale but also introduces variation issues which limits their practical use in accelerators.
These variations can occur between devices (device-to-device or spatial variation) or over time (cycle-to-cycle or temporal variation), ultimately restricting the number of states that can be stored and computed in the device.
The inaccurate programming issue caused by device variation can be mitigated through an iterative program-and-verify process. However, this approach results in high write energy and latency. Despite this, these drawbacks are generally acceptable since they represent a one-time overhead for inference accelerators.
Research has been reported to reduce the programming variability by engineering the devices, such as using 1D channels in epitaxial dielectrics\cite{choi2018sige}, confinement nanolayers \cite{ding2019phase}, and introducing an annihilation or completion voltage in the denoising process to the incomplete channels in memristors, enabling thousands of device levels \cite{rao2023thousands}. 

Device drift issues still cause performance degradation over a long time. 
Conductance state stability is crucial for applications that demands a stable synaptic connections in artificial neural networks for inference. 
Increasing the potential barrier between switchable states is a way to improve the stability and retention of the conductance state, but also increases the switching energy. Works have shown that using TaOx conduction channels can achieve a retention of $\geq$10 years at room temperature\cite{jiang2016sub}.
On the other hand, conductance decay can also be harnessed to simulate specific synaptic and neural dynamic behaviors and generate random numbers\cite{wang2017memristors}. 

\subsubsection*{\textbullet{High conductance states}}
    
Memristors based on filaments typically have high conductance (low resistance), resulting in large currents during computation. 
This can be a big challenge because the high conductance in a large array of memristors causes higher power consumption and a more noticeable IR drop along the wires, which can lower the accuracy of analog computing.
On the other hand, an extremely lower conductance is also not favorable because that might impose higher requirements on the sensing circuits, for example, finer TIAs and longer integration time to accumulate the current, during both computing and read-and-verify processes.
Memristors in low conductance also leads to current-voltage (I-V) non-linearity (and thus lower computing accuracy) because of non-Ohmic conduction mechanism, and usually increases the programming voltage / energy.


The conductance range of reported memristive crossbar arrays normally spans from 10 mS to 1 $\mu$S (100 $\Omega$ to 1 M$\Omega$). 
Given that low resistance can cause various issues as discussed earlier, and since a dynamic range of 100 is generally sufficient for computing purposes, most recent studies use around 1-100 $\mu$S as the working range for in-memory computing in crossbars. In contrast, a wider range is used for in-memory searching in CAMs.
It is generally believed that the ON-state current does not scale for redox memristors due to the localized conductance mechanism, but the OFF-state current can be reduced because of smaller leakage. 
Tantalum oxide memristors with 25 nm lateral sizes, integrated on CMOS transistor circuits, have shown a six order-of-magnitude operating range, mostly due to the small leakage in OFF states. This allows for 3-bits conductance levels below 10 µS for low current operation. \cite{sheng2019AEM}. 
Lower current has been demonstrated in HfO2/TiOx crossbars with a half-pitch of 6 nm and a critical dimension of 2 nm\cite{pi2019memristor}. 
Embedding nano-structures in the device can also impact the conductance range. For example, studies have shown that adding nano-particles to HfOx \cite{wang2018interface} or introducing metal nano-islands in the HfOx layer with different metals \cite{wang2019highly}
The conductance range can be improved via various methods such as doping of resistive switching layer \cite{sedghi2017role}\cite{li2019improvement} or top electrode \cite{li2020indirect}, and bulk switching memristors\cite{wu2023bulk}.



\subsubsection*{\textbullet{Write endurance, energy and linearity}}
    
Most memristor-based machine learning accelerators target workloads that avoid frequent memristor updates because of high write energy and limited endurance. 
However, there's still a need to accelerate workloads with frequent weight updates, such as KVQ updates in the multi-head attention module of Transformer model inference and on-chip training\cite{yang2013memristive} of all neural network models. 
In these cases, limited write endurance and high write energy present significant challenges.

Endurance failure of devices may originate from structural fatigue, particularly in redox and phase-change materials. For redox materials, these fatigues include undesired redox reactions with electrodes, overgrowth of filaments, or unwanted diffusion (or loss) of filament atoms \cite{wang2020resistive}. A thermodynamically stable redox resistive switching material can achieve an endurance of more than $10^7$ cycles \cite{yang2010high} or $10^{12}$ cycles \cite{kempen202150x}. In the realm of phase-change materials, endurance strongly correlates with electromigration and the mechanical stress induced by variations in density.\cite{raoux2008phase}. The endurance of phase-change memristors can be increased by engineering the film-deposition technique, composition, and device geometry\cite{padilla2011voltage}. Previous work has shown the endurance of a GST PCM device was measured up to $10^8$ cycles\cite{ji2018IEDM}.


Another challenge in device states writing is ensuring the linearity and symmetry of a memristor's response to programming pulses. 
This is crucial for accurate weight updates, especially during on-chip neural network training. 
Without these features, an auxiliary memory would be needed to track the current state and calculate the next programming pulse, or one would have to use costly iterative program-and-verify methods, both of which are too expensive for efficient on-chip learning.
Many device efforts that reduce variation also improve update linearity and symmetry, such as the previously discussed method using threading dislocations in SiGe to confine metal filaments in a single-crystalline SiGe layer epitaxially grown on Si\cite{choi2018sige}.
Additionally, work has also shown that homogenizing the filament growth/dissolution rate via the introduction of an ion diffusion limiting layer (DLL) at the TiN/TaOx interface can strongly increase the linearity of conductance modulation\cite{wang2016engineering}. 
Electrochemical random access memory (ECRAM) is another promising device type that decouples the read and write paths, demonstrating promising conductance update linearity and symmetry\cite{talin2023ecram}\cite{chen2023open}.

Continued efforts to improve the write energy, endurance and linearity are still needed to enable workload acceleration that demands frequent updates.

\subsubsection*{\textbullet{Future directions}}
\begin{figure}[H]
    \centering
    \includegraphics[width=1\textwidth]{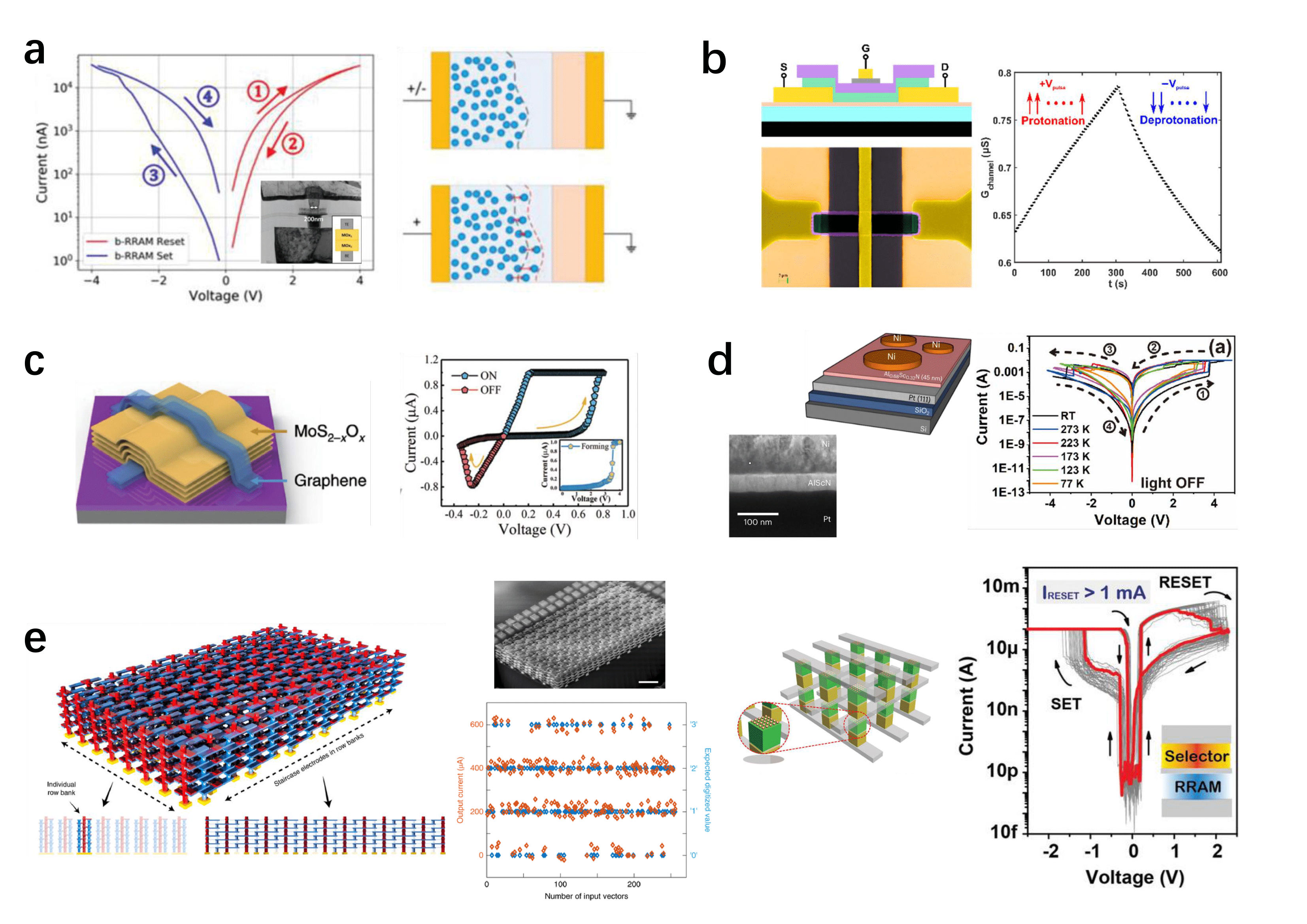}
    \caption{Future directions for devices. (a)Bulk switching mechanisms, adapted from \cite{wu2023bulk} (b)ECRAM, adapted from \cite{talin2023ecram} (c)2D materials, adapted from \cite{Mengjiao2023}  (d)Extreme condition, adapted from \cite{Pradhan2024}\cite{RANI2022174} (e)3D stacking, adapted from \cite{lin2020three}} 
    \label{fig:future direction}
\end{figure}
\begin{itemize}
    \item Exploring alternative device material stacks and switching mechanisms

   Admittedly, memristors, especially the commercially adopted ones, face challenges. Future device development and optimization are necessary, but this might involve creating different types of switching mechanisms and material stacks. It's important to note that optimizing one specific performance metric often comes at the expense of others. A universal memory that outperforms all others in every metric is unlikely due to physical mechanism constraints.

   Recently, there have been reports exploring devices with bulk switching mechanisms\cite{wu2023bulk}, demonstrating improved uniformity in large devices, though scalability down to several nanometers remains unclear yet. 
   ECRAM is another example that adopts a new switching mechanism using electrochemically inserted ions to modulate the conductance between the other two terminals\cite{talin2023ecram}\cite{chen2023open}. 
   This new mechanism also offers a different operation procedure, with separated write and read/compute paths. While the write energy and linearity have shown significant improvement, data retention, especially at higher temperatures, remains a concern.

    Two-dimensional (2D) materials could bring new physics to memristor devices, resulting in various performance trade-offs. High-quality 2D material can provide atom-level uniformity and is promising to effectively reduce the D2D and C2C variation\cite{Mengjiao2023}, which are common in memristors based on bulk materials. The structural defects and compositional imperfections in 2D materials are highly tunable by external field, and thus can provide richness in methods to design the switching of memristors\cite{Mengjiao2023}. However, achieving large-scale integration of 2D materials with industry-standard uniformity remains a significant challenge, and requires a collective efforts at material, process and device levels. 
    A memristor crossbar array is demonstrated using wafer-scale polycrystalline HfSe2 grown by molecular beam epitaxy and a metal-assisted van der Waals transfer technique \cite{LiAM2021}. Another work shows that high-density and yield memristive crossbar arrays can be fabricated using CVD-grown h-BN as the resistive switching material \cite{Chen2020NE}. 
    Solution-processed two-dimensional MoS2 memristor arrays are reported with excellent endurance, long memory retention, low device variations, and high analog on/off ratio with linear conductance update characteristics. \cite{tang2022wafer}.

    \item Improving density with three-dimensional stacking

    Three-dimensional stackability is the key strength of memristor technology because the process is back-end compatible and subsequent process for upper layer usually have little impact on underling layers.
    This is a key strength for an accelerator targeting running large models. 
    There have been demonstrations of three-dimensional memristor arrays and computing accelerator prototypes. For example, some works have performed dot-product operations with 3D memristive crossbars to show the applicability of such 3D CMOL hybrid circuits \cite{Payvand2015ISCAS}. A 3D circuit consisting of eight layers of monolithically integrated memristor devices has also been reported, implementing a CNN function by programming parallel operation kernels into a 3D array. This achieved software-comparable accuracy (98.1\%) in recognizing handwritten digits in an the MNIST dataset \cite{lin2020three}\cite{hua2019advs}.

    One challenge in memristor 3D stacking is the backend selectors. Front-end transistors are usually used as selectors for one-transistor one-memristor (1T1R) configurations, or other variations (2T2R, 1T2R, etc.) in recent prototypes. However, these selectors also limit the 3D stackability. 
    Efforts have been made to develop back-end transistors, such as thin-film oxide transistors (e.g., IGZO transistors)\cite{Cho2021IGZO} or 2D material-based transistors \cite{Wang20242DT}. 
    Another solution is to develop back-end selector devices, which have been used by Sandisk\cite{liu2013130} and Intel/Micron\cite{Merritt20163d} for memory products, or to develop self-selecting devices.
    Current reports are showing that self-rectifying devices can suppress the leakage current to an unnoticeable range and can achieve a read margin to the order of 100$\times$100 array size or even larger\cite{Lim2024acsnano}. However, current self-rectifying devices suffers from large operating voltage and high non-linearity\cite{Minjae2024AMI}. Developments should be made in this direction for a more efficient accelerator array.

    Heat dissipation is another challenge in creating ultra-high density, 3D stacked memory. Higher thermal density and the interlayer dielectric, which is usually less thermally conductive than the silicon substrate, add to the difficulty. Addressing this issue requires collective efforts in developing more thermally conductive insulating materials, multi-physics simulations, and workload distribution. This can be an interesting research direction.

    \item Demonstrating the operation under extreme conditions

    Due to its working principle, the memristor device has the potential to deliver performance that conventional charge-based solid-state memory cannot, such as operating under extreme conditions like extremely low or high temperatures or highly radioactive environments\cite{RANI2022174}\cite{Pradhan2024}. 
    There is significant demand for these capabilities. 
    For instance, memory for cryogenic quantum computing peripherals may need to function at low temperatures, while aerospace applications may require electronics that can operate in high radiation and high-temperature environments. 
    Validating operations under these conditions and optimizing performance for them could represent a future direction.

\end{itemize}

\subsection{Circuit}

Having discussed the device-level challenges, it is equally important to consider the circuit-level aspects of memristor-based machine learning accelerators. The performance and efficiency of these accelerators heavily depend on the circuits designs, including input encoding and driving circuits, crossbar array designs, output sensing circuit techniques. However, designing these circuits presents its own set of challenges due to the unique characteristics of memristors and the requirements of machine learning workloads, which could be significantly different from the designs of those components for other purposes. In this section, we will share our opinion on those key circuit-level challenges that need to be addressed to fully unlock the potential of memristor-based accelerators. 

\subsubsection*{\textbullet{Input driving circuit}}

The first challenge is how to represent the input and apply it to the crossbar array. The most natural idea is to use voltage amplitude for representing the input and directly apply it to each row (bit line) of the crossbar array\cite{li2017analogue}. The advantage of this implementation is that the computing can be done in a single time step causing ultra low latency. And it can be potentially used for processing the analog single generated by sensors without the need of data conversion. However, for mixed-signal system, this implementation requires a large number of high-resolution DACs resulting in huge power and silicon area consumption making it challenging to be fully integrated on chip. Moreover, encoding input as voltage amplitudes imposes higher requirement on the I-V linearity of the memristor device. For filamentary RRAM devices, better I-V linearity can usually be achieved by setting them to to the high conductance range, as Ohmic conducting mechanism dominates. 
However, the non-linearity problem becomes considerable at the low conductance range due to different conducting mechanism and is more severe considering that design nowadays favors lower conductance for improved energy efficiency.

While this issue can be potentially be tackled by device efforts, as introduced in the earlier section, one possible circuit solution for this is to use pulse width  \cite{cai2019fully}\cite{khaddam2021hermes}\cite{khaddam2022hermes}\cite{le202364core}\cite{wan2022compute} (i.e. using pulse width to represent the input) for encoding input. In this way, the voltage amplitude can be fixed, and thus can mitigate the I-V linearity problem. Moreover, the pulse width modulation (PWM) circuit can be designed quite compact enabling it to be connected to each row of the array and realizing full parallelism. 
However, this method requires additional accumulation circuits at the output and introduces $2^N\times$ latency overhead, which largely sacrifices the throughput. 

Another compromising solution is input bit slicing, i.e. for an N-bit input, to apply N binary voltages from LSB to MSB to the array in N time steps and conduct the multiplication for N times\cite{li2020cmos}\cite{correll20228bit}. This method also only needs a 1-bit DAC for each row, which can be implemented by a analogue multiplexer multiplexing between the two DAC output states and can reduce the latency overhead to N times, however it either needs a complex analogue circuit to accumulate the result \cite{correll20228bit} or needs to do the analogue-to-digital conversion for N times, followed by an digital shift-and-add operation block. 
All these factors might reduce the energy efficiency of the computing.
Furthermore, for the bit slicing, it does not necessarily need to be sliced to binary states. For example, Ref. \cite{xue202122nm} splits the 8b input to 2b, 3b ,3b to prioritize the MSB accuracy and reduce the latency overhead brought by a complete binary bit slicing. And the above mentioned methods are not mutually exclusive, but can be combined to take the advantage of both. For example, in Ref. \cite{wan2022compute}, a 6b input is split into two 3b inputs with each represented by pulse numbers. 

One challenge for those implementations using low-precision DACs and multiplexing between the DAC output states is that one DAC output channel might need to drive multiple rows of the array. Thus, it is required to have a low output impedance especially for a large array. One method to mitigate this is to limit the row number that can be activated each time step. However, this will affect the parallelism and thus the computing throughput. Another way is to use additional amplifiers as the output buffer of the DAC, which introduces extra area and energy consumption. For those cases that unitize binary input states, it is possible to connect the input to the word line of the array (i.e. the gate terminal of the accessing transistor in a 1T1R structure)\cite{hung2021four}\cite{jiang202240nm}, so that the DAC does not have a driving issue. However, additional driving circuits are still needed on the bit line to enable enough current flowing through the array. 

We would believe that a perfect solution for the input circuit does not exist. Aiming at different application requirements, balancing well the trade-offs between accuracy, latency, area and energy consumption is another challenge on the design site. 

\subsubsection*{\textbullet{Crossbar array}}

Crossbar is the structure that performs in-memory multiplication, so the design mainly focuses on: 1. ensuring the accurate programming of conductance values representing the matrix, and 2. ensuring accurate computation with the programmed conductance values representing the matrix. To achieve accurate programming, inter-cell interference caused by the ``sneak path'' issue must be addressed. This is typically done by adding an address selection selector, usually an oversized transistor, to pass the large current needed for the reset operation, which compromises memory density. While density is critical for memory applications, it is usually acceptable for computing purposes. Further improvement remains a challenge that requires collective efforts at both the device and circuit levels.

As the crossbar array scales up, the calculated matrix increasingly deviates from the programmed value due to parasitic resistance and capacitance. 
Parasitic resistance causes the actual voltage drop on the memristor cell to differ from the applied voltage on the crossbar row because of the IR drop on the wires (both rows and columns), reducing computing accuracy. 
This issue worsens with larger array sizes, limiting the crossbar array's scalability.
 
There are several methods to mitigate the problem. 
One approach is to change the routing to minimize the effect. The most straight forward method is to connect the input and output to the array with more wires\cite{chen2013comprehensive}, reducing the average distance between the accurate input source and the actual memristor cell and thus the IR drop. This can be done by connecting both sides of a row to the same input and connecting both sides of a column to the same output, or by inserting some nodes inside the array and connecting them together\cite{shang2020fast}. However, the complex routing in the layout design can be challenging as each wires connecting the nodes also need to be wide enough to minimumize the effect. Another method is to construct a differential 2T2R cell with two inputs of the same amplitude but different polarity. In this structure, current subtraction occurs within the cell, which reduces the current on the source line and thus alleviates the IR drop effect\cite{liu2020fully}. However, this method can only partly alleviate the problem on the source line and is not helpful for the IR drop effect on the bit line.
 
Another mitigation method is to adjust the memristor conductance to counteract the effect. 
The system remains linear, so the crossbar still performs a linear transformation on the input vector. However, this transformation is not exactly the conductance matrix programmed into the memristors due to the IR drop issue. Knowing the matrix we want to multiply, we can calculate the necessary memristor conductance using the wire resistance information. 
It is usually done by first estimating the effect of IR drop, and then calculating the conductance value needed to compensate the effect and programming the calculated conductance value to the array so that the effective conductance after considering IR drop effect can match the target weight\cite{liu2014reduction}\cite{feinberg2021analog}\cite{gao2022fast}. 
However, the effectiveness of this method might be harmed by limited device precision, as the method reduces the mapped conductance range and thus increases the requirement for device precision in some parts of the crossbar, especially in a large array. Moreover, it suffers from device I-V non-linearity, the effect of which cannot be easily modeled and compensated. 
 
In addition to these methods, one method is to add additional calibration steps on the MVM output to recover the computing accuracy affected by IR drop in some extent\cite{li2017analogue}\cite{khaddam2021hermes}\cite{khaddam2022hermes}. A linear correction process is usually used for the calibration. In initialization phase of the linear correction, several random input are sent into the crossbar array. And a linear fitness between the experimental output and expected output is done for each output channel. In subsequent usage, this linear fitness function is applied to the output to partly recover the computing accuracy. 
This linear correction can be efficiently implemented in hardware by fine tuning the sensing circuit. However, this method cannot fully address the problem and introduces time, area and energy overhead on the calibration circuits. 
    
As for the parasitic capacitance, it majorly affects the settling time of the computing and thus influences the throughput. If PWM is used as input, large parasitic capacitance also influences the computing accuracy as the input signal got different levels of distortion at different location of the array. Reducing the parasitic capacitance mainly relies on careful layout design, which is another challenge on the design space.

\subsubsection*{\textbullet{Output sensing circuit}}
Similarly with input driving circuits, output sensing circuit design is another challenge at the macro circuit level. The output sensing procedure is usually composed of two parts: collecting output and analog-to-digital conversion. 

The output collection circuit is also determined by how the crossbar performs the computation. 
Early works mostly use current sensing, and in this method, the bit line is clamped to ground. This ensures the voltage drop on the memristor cell equals to the input voltage, making the current flow through each memristor cell to be the scalar multiplication of input voltage and the memristor conductance.
In this case, the bit line current or integrated charge represents the multiply-and-add (MAC) result. 
One implementation of collecting output of current sensing is to use a trans-impedance amplifier (TIA) or an integrator to both clamp the bit line voltage and covert the current to voltage so that it can be processed by normal voltage-based ADCs\cite{li2020cmos}. 
One problem for this implementation is the large energy and area consumption on the TIA or integrator. Additionally, the settling time of the TIA or integrator might also influence the computing latency. Another implementation method is to use a voltage regulator to clamp the bit line voltage and directly use a current-based ADC to covert the current to digital output\cite{khaddam2021hermes}. However, the design difficulty for current-based ADC is much higher than voltage-based ADC. Furthermore, a common problem for current sensing is that amplifiers usually suffer from low gain at advanced technology node, which causes the bit line voltage cannot be perfectly clamped to zero and thus harm the computing accuracy\cite{chakraborty2020resistive}. 
Another issue with this method is that memristor conductance tends to be quite high, leading to very high collected current in a dense matrix within a large array. This presents significant challenges for the driving circuit and the overall power consumption of the circuit macro.

An alternative approach is to use voltage sensing. Here, the bit line current is kept at zero, and the multiply-and-add (MAC) result is represented by the voltage on the bit line, which is usually sampled by a simple capacitor connected to the bit line\cite{wan2022compute}.
This method saves the consumption on the TIA or voltage regulators, and makes it easier to keep the bit line current at zero with a high impedance device. 
However, the sampled voltage is the MAC result normalized by the sum of the conductance of the cells in the same column. 
This normalization factor usually needs to be calculated first and calibrated back in the digital domain, adding overhead. 
Alternatively, a differential pair on the same bit line can be used to address this problem. The difference in conductance between two memristor cells represents the weight, and the sum of the conductance of two memristor cell is set to a constant. This way, the normalizing factors of all bit lines become the same, making it a universe scaling factor and avoids the calibration process. 
However, the conductance variation of memristor can be a problem for this method, as the sum of conductance of one bit line cannot perfectly match other bit lines. Moreover, in this method, the conductance states are generally higher than other differential pair implementations and thus compromise the energy efficiency.

For anaolg-to-digital conversion process, usually a flash-ADC\cite{yin2019monolithically} or SAR-ADC can be used. A flash-ADC has lower latency but consumes larger silicon area and energy and cannot be designed to have very high resolution due to the effect of noises. A SAR-ADC has N$\times$ latency for N-bit resolution, but can be designed to have higher resolution. For both types of ADCs, if high-resolution is needed, it consumes too much silicon area to be accommodated one for each column. One compromising solution is to have several columns being multiplexed for one ADC at the cost of lower parallelism and lower throughput. However, the sampling frequency of state-of-the-art ADCs are much faster than the working frequency of the memristor array, one can use some sample-and-hold circuits to first hold the MAC result and wait it to be time-multiplexed processed by the ADC to minimize the latency loss. 

Moreover, recent paper explored using ramp-ADC for the conversion\cite{liu2020fully}\cite{ambrogio2023analog}. Although ramp-ADC needs $2^N\times$ cycles for the conversion, but the generated ramp reference can be shared across the entire tile. Thus, each column only needs a comparator and a counter, which is energy and area saving. And some simple activation function, such as Relu can be integrated in the ramp generation. Moreover, it is not always needed to completely convert the analog signal to digital signal. This ramp-ADC can stop at converting the analog amplitude value to PWM signal, which can be used as the input of next layer.

As for current-based ADC, Ref. \cite{khaddam2021hermes} implements a CCO-based ADC, which first converts the current signal to frequency and then uses a counter to covert it to digital output. This CCO-based ADC can be compact enough to match the number of columns, however suffers from high latency problem and low linearity.

For the case that PWM is used as input or input bit slicing is conducted, except from output collecting and analog-to-digital conversion, one additional circuit is needed to accumulate the result. This can either be done in digital domain, such as using a digital counter\cite{khaddam2021hermes} for PWM input or a digital shift-and-add circuit\cite{li2020cmos} for input bit slicing, or in analog domain, such as using a integrator to accumulate the charge\cite{wan2022compute} or using current mirror with different multiplier for the shift and add\cite{cai2019fully}. However, all these implementations will inevitable introduce additional overhead.

Although there are various methods to implement the output sensing circuits and the ADCs, the area and power consumption of the ADC part in a memristor-based VMM accelerator macro still dominates. Further optimizing the circuits according to different application requirements is challenging but necessary.

\subsubsection*{\textbullet{Programming circuits}}
Designing circuits to precisely and efficiently program the memristor cell is another challenge for the macro design. For training tasks, adjusting the weight stored in the memristor usually dominates the latency and energy consumption, due to the serial nature of programming scheme. 
For inference-only tasks, although it seems that the programming process is one-time overhead and can be acceptable, due to the retention and relaxation issue of memristor device, the computing error grows larger as time increases. Periodically refreshing the memristor conductance value might be necessary\cite{xiang2019impacts}, before the device performance can be further improved. 
For inference with Transformer-based models, such as large language models, there are parts that need to be frequently updates. 
All of these poses requirement on efficient programming circuits.
Another problem for the programming circuits is that advanced technology node can only support low VDD, lower than programming memristor devices. 
Charge-pump circuit and special transistors that can endure higher voltage needs to be used for designing programming circuits and the signal path, which increases the design complexity and usually consumes larger area and energy. 
Moreover, precise programming of memristor conductance relies on write-and-verify scheme. However, applying such scheme by external equipment is time-consuming. Several paper have demonstrated on-chip write-and-verify circuits\cite{li202240nm}\cite{yoon202240nma}\cite{yoon202240nmb}, however, it only supports programming of binary or a few conductance states. In order to fully unleash the analog conductance states potential of memristor device, higher precision and more efficient programming circuits is needed.

\subsubsection*{\textbullet{Future directions}}
Although there are various challenges remaining on peripheral circuit design. In the mean while, there are also opportunities at the circuit level to address problems or leverage properties of memristor devices.

\begin{figure}[H]
    \centering
    \includegraphics[width=1\textwidth]{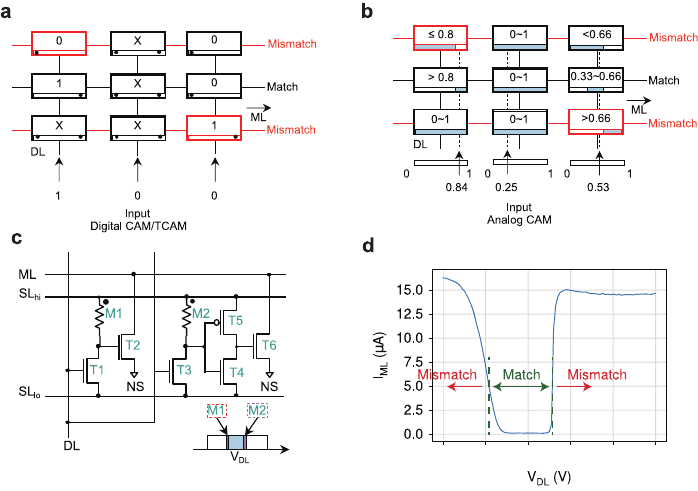}
    \caption{Content addressable memory (CAM) as a potential structure for machine learning. (a)Working principle of digital CAM or ternary CAM (TCAM). (b)Working principle of analog CAM. (c)Circuit structure of a single analog CAM cell. Two memristors are used to define left and right boundaries, respectively. (d)Match line current versus data line voltage. Panel a,b,c are adapted from \cite{pedretti2021treebased}. Panel d is adapted from \cite{li2020analogcam}.} 
    \label{fig:CAM}
\end{figure}

\begin{itemize}


\item Utilizing device non-ideality

Despite pursuing of improving the computing accuracy, some memristor devices can be used as the noise source for various random computing situations with special designed circuit. 
One typical application is to utilize the noise from the device for Boltzmann sampling\cite{shin2018hardware}\cite{mahmoodi2019versatile}\cite{yan2021reconfigurable}, which can be further used in simulated annealing to solve NP-hard problems or in Boltzmann machine for unsupervised learning, as seen in \ref{fig:probability-computing}(a)-(c). For this sampling, the key is the tunable sigmoid-like stochastic activation, which can be realized by stochastic diffusive memristor\cite{shin2018hardware}\cite{yan2021reconfigurable} or utilizing the intrinsic noise of memristor crossbar array\cite{mahmoodi2019versatile}. This eliminates the need for random number generator and the look-up table and thus largely increases the efficiency. Moreover, the same idea can be easily extended to deep belief networks for more complex tasks\cite{wang2022memristive}. In addition to Boltzmann sampling, the noisy property of memristor device can help to largely reduce the cost for implementing Bayesian neural network\cite{li2022enabling}\cite{lin2023uncertainty}, which usually has large overhead when implemented with conventional hardware,as seen in \ref{fig:probability-computing}(d)-(f). One promising direction on the circuit level is to design specialized circuits according to the device property to fully unleash the power of memristor-based analog computing.

\begin{figure}[H]
    \centering
    \includegraphics[width=1\textwidth]{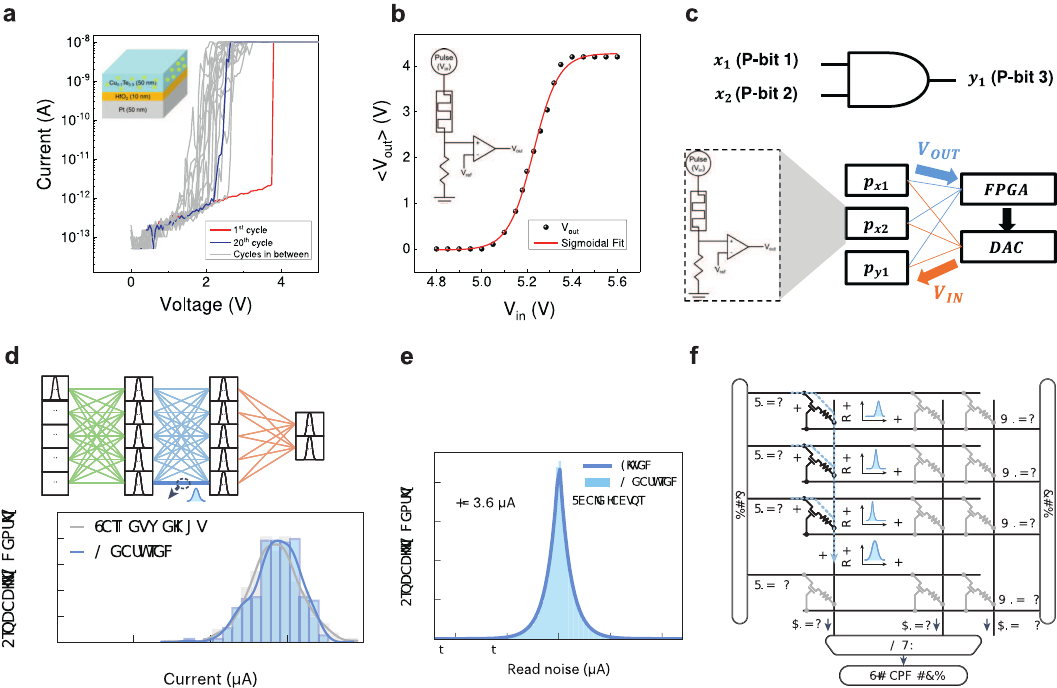}
    \caption{The noises of memristor device can be utilized for probabilistic computing. (a)Stochastic switching of diffusive memristor (b)Sigmiod-like probability activation generated by the circuit (c)Hardware implementation of stochastic AND gate for probability computing (d)The structure of the Bayesian neural network. The distribution of the weight can be implemented by the distribution of the memristor conductance. (e)Read noise of memristor (f)Circuit realization of controllable Gaussian distribution needed for Bayesian neural network. Panel a,b,c are adapted from \cite{woo2022probabilistic}. Panel d,e,f are adapted from \cite{lin2023uncertainty}.} 
    \label{fig:probability-computing}
\end{figure}

\item Non-crossbar structure

In addition to crossbar-based VMM accelerator, special circuits can be designed for different structures for efficiently implementing various ML tasks. Content-addressable memory (CAM) is one of the promising ones among various structures, as seen in \ref{fig:CAM}. Researchers have explored of utilizing CAM for accelerating explainable AI models such as tree-based models\cite{pedretti2021treebased}. Also, memristor-based CAM can be used as the external memory needed in memory augmented neural network(MANN)\cite{mao2022experimentally}, enabling the network to retain and access information over time, which potentially gives the network ability of processing long-term sequences and large-scale data and also the ability of learning from limited number of training samples. Designing circuits needed for CAM might differ a lot from designing circuits for crossbar array, but we believe that it is worth exploring under this direction.
\end{itemize}

\subsection{System}
\graphicspath{{figures/}}
System design as the bridge between device/circuit and algorithm/application should take into consideration of the difficulties from both upper and lower abstraction levels. Therefore, careful co-design and co-optimization across various layers is a major challenge for the system design with emerging non-volatile memristor devices.

\subsubsection*{\textbullet{Co-design for device non-idealities}}

First, the system design for memristor-based machine learning accelerators should take into careful consideration the device non-idealities, based on how many device challenges are addressed.
Different design techniques have been proposed to mitigate the negative impact of device non-idealities.
For example, the bit-slicing technique has been utilized to design the system with limited single-device bit accuracy \cite{ankit2019puma, ankit2020panther}. Under this scheme, the synapse weight is encoded into eight memristors, each with 2-bit accuracy, which can be achieved reliably in memristor devices. 
The method has a limitation: while these devices can achieve 2-bit information storage when used as memory, they pose a challenge for computing purposes. This is because the device conductance in computing is read as an analog value instead of a quantized state. 
A recent breakthrough involves the implementation of adaptive analog slicing, based on the programming accuracy of the more significant bit, achieving an arbitrary equivalent precision of the device programming \cite{Wenhao2024Sci}. However, careful optimization is still needed to trade-off computing accuracy, machine learning algorithm requirement, and the system power/area overhead. 


Despite these efforts, defective devices may still appear due to limited fabrication yield or endurance performance, deteriorating classification accuracy. Verification methods are crucial to ensure system health and computing fidelity, distributing the workload to healthy elements or re-programming and re-calibrating defective ones. 
One approach is analog error correcting codes (ECC), which verify results during the computing process without significant overhead and offer the possibility to correct results without re-calibrating the system \cite{li2020analogecc}. 
Careful co-design is needed when designing system with analog ECC because there are trade-offs between the guaranteed accuracy and additional overhead, computing block reprogramming overhead, and error correction overhead, etc. 


\begin{figure}[h]
    \centering
    \includegraphics[width=1\textwidth]{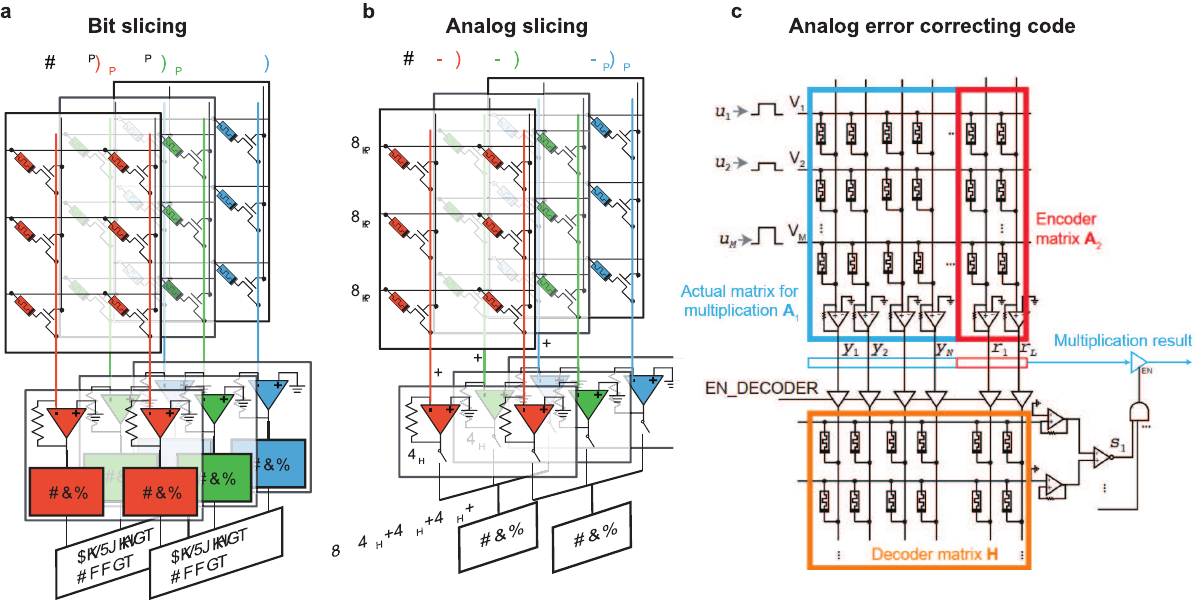}
    \caption{Software hardware co-optimization for improving computing accuracy and reliability. (a)Digital bit slicing (b)Analog slicing (c)Analog error correction codes Panel a,b are adapted from\cite{Wenhao2024Sci}. Panel c is adapted from \cite{li2020analogecc}.} 
    \label{fig:codesign-variation}
\end{figure}

\subsubsection*{\textbullet{Co-design for circuitry trade-offs}}

Besides the device challenge, circuitry also limits computing accuracy and needs careful design consideration. 
There are multiple trade-offs between computing accuracy and chip area, energy, and speed overhead. 
On one hand, we should custom design the peripheral circuitry, especially for the major consumer, in particular the analog-to-digital signal conversion circuitry, to meet specific requirements like low accuracy or prioritizing energy efficiency over speed.
\cite{khaddam2021hermes, khaddam2022hermes, zhang2023edge, yao2020fully, cai2019fully}. 
On the other hand, algorithms can be custom developed to better tolerate the lower input/output bit precision and synapse weight accuracy that are stored in memristor devices without signficantly sacrificing model performance , such as recognition accuracy.
For example, it has demonstrated binary neural network (BNN) can achieve an acceptable recognition accuracy while offering significant benefits in terms of area and energy consumption, as well as greater tolerance to device variations when deployed in memristor-based hardware.\cite{bocquet2018memory, yin2020high}.
To fully utilize the analog conductance states in memristor devices, defect-aware training has been proposed to train a model, taking into consideration device statistical variations and limited accuracy caused by the peripheral circuit design before deploying them to hardware \cite{Mao-2022-Experimentally-Validated}. 
Spiking neural network (SNN) has also been proposed because of its advantages in encoding information in time domain and thus doesn't require ADC \cite{yin2023accurate, Akopyan-2015-TrueNorth}. However, SNN still suffers from relatively low recognition accuracy. It's ability to scale up to deep neural network and different forms of neural network such as CNN is unknown.
A comprehensive system solution is a complicated multi-object optimization problem, involving multiple trade-offs between model performance and peripheral circuit overhead given device performance constraints with numerous parameters. 
Neural system search can be explored to tackle this challenge, but an optimal practice is still an open question.



Great progress has been made in the recent five years in system level to demonstrate a whole system on chip (SoC) as the neural network accelerators \cite{xue20191mb, su2017462gops, yao2020fully, ambrogio2023analog, chen2019CMOS, xue2020CMOS, hung2021four, le202364core, zhang2023edge}. 
These accelerators promise tremendous improvement in terms of speed, energy efficiency and chip area efficiency compared with traditional CPU/GPU, but there still remain challenges. 
Firstly, current developments mainly focus on accelerating the matrix multiplication operation. However, other operations, such as normalization, activation, and pooling (which usually have a time complexity of $O_(n)$), as well as input data preparation for CNNs for offline inference, still incur significant overhead.
For online training, vector gradient calculation and outer product accumulation (OPA) are also needed. 
Most of these operations are currently implemented in the digital domain, making the high-bit ADC the performance bottleneck. Attempts have been made to implement activation in analog circuits \cite{kian2021fully} and OPA in-situ in memristor crossbars \cite{ankit2020panther}\can{wrong citation?}. 
Efficiently implementing these functions in hardware (analog or in-memory) and integrating them with memristor crossbar vector matrix multiplication (VMM) remains an interesting challenge. 
Another problem is controlling the data flow and communication between tiles. Due to the parasitic resistance and capacitance, the size of the computing core, memristor crossbar, is limited up to several hundreds to one thousand number of rows or columns. Thus, to handle larger data inputs and offering flexibility to various model size while maintaining array utlization rate, computations have to be divided into several memristor tiles. 
Attempts have been made to use switches-controlled all-to-all connected wires \cite{ambrogio2023analog,le202364core} and memristor crossbars as inter-tile connections \cite{Cong-2011-mrFPGA}. 
However, the optimal size of the crossbar in a tile and the optimal way to make inter-connection between tiles remain open questions. 


\can{Optional: Another direction to consider is thinking about system for monolithic and heterogeneous 3D IC.}

\can{(I think I don't have the knowledge to discuss about the ISA and compile part. Maybe it's better just not to discuss about it.) To decide whether we want to expand the software interface / ISA development: \\
For applications on edge side, with limited chip area and power supply, memristor based AI accelerators show superior performance because of its parallel data processing and non-volatile nature which leads to zero leaky power. Further research can be done to design ultra-lower power system and further reduce the operation voltage of memristor device to reduce the device programming power consumption. The interface with software programming is also significant. With the rapid development of AI technique, more and more algorithms and models will be proposed to put into practical application. 

To design an system to be compatible with the mainstream neural network models such as fully connected layers, CNN, RNN, and transformer together with a complete instruction set system (ISA) and compiler will be a very important research direction. }

\subsubsection*{\textbullet{Future directions}}
\can{Since the information in this paragraph is quite vague, let's consider moving the training and heterogeneous accelerator parts here as a potential future direction. You can incorporate some sentences of the writing to the corresponding sections}

\begin{itemize}
\item In-situ training and on-chip learning

Training is another seldom touched territory because it is more challenging to implement on memristor-based hardware. 
There remains clear motivation to implement and/or accelerate training on memristor-based hardware, because: 
1. Training a modern model, for example, a large language model, imposes significant energy challenges.
2. There is a need to adapt to device state drift and new environments.
Most prior discussions focus on inference acceleration, i.e., deploying an offline trained model to the chip and accelerating the neural network inference on-chip\cite{xue20191mb, su2017462gops, yao2020fully, ambrogio2023analog, chen2019CMOS, xue2020CMOS, hung2021four, le202364core, Dong-2019-Convolutional}.
However, implementation of training on-chip faces challenges across different levels: device endurance and variation, back-propagation of errors, and additional memory needed to store activation values and matrix gradient for batch training. 

First, for online training, synapse weights needs to be updated after each batch, but this frequent updating is constrained by the memristor device's limited writing endurance and high write energy and speed overhead. 
An endurance-aware training method \cite{Yang-2023-ESSENCE} has been proposed to mitigate this issue by reducing the number of write operations, while slightly sacrificing the classification accuracy. 
Second, in the most prevalent training algorithm: back-propagation (BP), while while the error back-propagation can be efficiently implemented in the crossbar, the forward propagation and back-propagation need to be done sequentially  due to time dependence, which prevents layer-wise parallelism in crossbars.
What's worse is that calculating the vector gradient and outer product imposes even larger latency, energy and chip area overhead than the crossbar itself.
The needs of storing activation values (especially for recurrent neural networks) for all layers as well as the matrix gradient add additional memory unit, which will cause additional data movement between the memory and computing unit and will become the performance bottleneck. 
Bio-inspired local training algorithms and principles such as spike time/rate dependent plasticity (STDP) \cite{Huang-2019-Binary, wang2017memristors} have also been proposed as local synapse updating rules that are easier to be implemented in hardware for online training. But these bio-inspired weight updating rules lack theoretical support for high model performance such as classification accuracy. 
Due to these challenes, while offline inference hardware implementation can already deal with relatively complicated classification task such as Cifar-10 dataset but fully online training still remains challenging. 

As discussed above, BP-free algorithm is preferred in hardware implementation. Among them, the Forward-Forward algorithm \cite{hintonForwardForwardAlgorithmPreliminary2022, renScalingForwardGradient2023} is one of the most promising and hardware-friendly BP-free algorithms because it combines the advantages of Local Greedy and Forward Gradient algorithm and has the best performance scaling up to a 58.37$\%$ testing error on ImageNet dataset \cite{renScalingForwardGradient2023}. Forward-Forward algorithm also doesn't require storing the activation values which alleviate additional memory needed for online training. From application perspective, memristor-based AI accelerator is more suitable in edge computing. Thus, future design should focus more on ultra-lower power and system reconfigurability, to be compatible with different AI models. 
In order to achieve this, a comprehensive modelling at different levels is required for application-specific co-design.


\item Heterogeneous accelerators with different memory technologies

Despite efforts in memristor material and device optimization, it is widely acknowledged that machine learning accelerator designs based on current memristor technology should avoid frequent updates. Consequently, memristor-based accelerators are more promising for edge devices, where the computation is intermittent and extremely low energy consumption is required. In this scenario, the non-volatility of the device can be fully exploited to achieve zero static power consumption.
However, the technology is currently not suitable for accelerating workloads in data centers due to the heavy, constant computing demands with massive amount of data input and output. 
Conventional computing system together with SRAM might be an option, because of its extremely write endurance and low update energy. However, in-memory computing accelerators based on SRAM still faces challenges in data centers, because of challenges in storage density and software compatibility.\cite{Lu-2024-High-speed}. 
Therefore designing an in-memory computing accelerator should carefully consider the application scenario and the features of the chosen memory technology.

Similar to memory hierarchy design, it might not be possible to find a single universal memory technology that fits all in-memory computing workloads. In this case, designs combining different memory technologies might be considered. This has received significant attention due to the rise of Transformer models and generative AI models, including large language models (LLM) based on them. Unlike conventional neural network inference, where all weights are stationary, models with attention mechanisms, such as Transformers, require frequent updates of $K$, $V$, and $Q$ matrices and computation with them. It is more practical to implement the parts that require frequent updates in volatile memories, such as SRAM, while keeping the stationary weights in non-volatile memristors. 
Combining different memory technologies also opens up new possibilities for on-chip continual learning and using volatile memories for frequent updates. 
A large space is opened in researching accelerators that combine different memory technologies, including process compatibility, inter-connection, and software resource allocation with new instruction set systems (ISA).

There are three different applications that are suitable to combine different memory technology or different in-memory computing paradigm. First, large language model (LLM) has shown tremendous success in pattern recognition, natural language processing, and logic inference. Although efforts has been made to accelerate transformer based model \cite{Yang-2022-Full-Circuit}, but the attention mechanism is still hard to be implemented in hardware. TCAM-based paradigm may accelerate part of the calculation in attention mechanism. Second, in few-shot learning and hyperdimensional computing, crossbar can be used to implement the feature extractor. LSH and TCAM can be used in the memory part for searching and recognition. Third, in multi-layer perception neural network, memristor technology and static random access memory (SRAM) technology can also be combined together to use memristor to represent the bits that are less frequently updated and to use SRAM to represent the bits that are more frequently updated \cite{Wen-2024-Fusion}. This new paradigm can preserve the advantages of both accuracy and energy efficiency \cite{Wen-2024-Fusion}.

\end{itemize}

\section{Summary}
In this paper, we have seen the prosperous development of memristor-based accelerators. And we have discussed about the critical challenges we are still facing at three different levels, device level, circuit level and architecture level. Although there are various challenges that are not easy to be handled easily, there are also various potential directions that can be great opportunities waiting for exploration. We believe that the future development of memristor-based accelerators relies on knowledge across all three levels. Focusing on one aspect usually cannot solve the problem thoroughly. Aiming at specific application requirement, balancing well the trade-offs and cooperating methods from different levels might be the key for net-step development.

\section*{Acknowledgement}

This work was supported in part by RGC (27210321, C1009-22GF, T45-701/22-R), in part by NSFC (62122005); and Croucher Foundation. 
Any opinions, findings, conclusions, or recommendations expressed in this publication do not reflect the views of the Government of the Hong Kong Special Administrative Region or the Innovation and Technology Commission

\label{}





\bibliographystyle{elsarticle-num} 
\bibliography{memML.bib}




\end{document}